\documentclass[%
 reprint,
showpacs,
 amsmath,amssymb,
 aps,
]{revtex4-1}

\usepackage{multirow,booktabs}
\usepackage[table]{xcolor}
\usepackage{graphicx}
\usepackage{dcolumn}
\usepackage{bm}
\usepackage[breaklinks]{hyperref} 
\usepackage[all]{hypcap}

\usepackage{relsize}
\newcommand{\ev}{{\rm e}\kern-1.pt{\rm V}}
\newcommand{\gev}{{\rm Ge}\kern-1.pt{\rm V}}
\newcommand{\mev}{{\rm Me}\kern-1.pt{\rm V}}
\newcommand{\kev}{{\rm ke}\kern-1.pt{\rm V}}
\newcommand{\tev}{{\rm Te}\kern-1.pt{\rm V}}
\newcommand{\gevsq}{\mbox{$\mathrm{{\rm Ge}\kern-1.pt{\rm V}}^2$}}

\def\cesrta{{C{\smaller[2]ESR}TA}}
\def\lsim{\mathrel{\rlap{\lower4pt\hbox{\hskip1pt$\sim$}}
    \raise2pt\hbox{$<$}}} 
\def\gsim{\mathrel{\rlap{\lower4pt\hbox{\hskip1pt$\sim$}}
    \raise2pt\hbox{$>$}}} 

\begin{document}
\title{
Measurement of Electron Trapping in the CESR Storage Ring}
\author{M.G.~Billing}
\author{J.~Conway}
\author{E.E.~Cowan}
\author{J.A.~Crittenden}
\email{crittenden@cornell.edu}
\author{W.~Hartung}
\author{J.~Lanzoni}
\author{Y.~Li}
\author{C.S.~Shill}
\author{J.P.~Sikora}
\author{K.G.~Sonnad}
\altaffiliation{Present address: KEK, 1-1 Oho, Tsukuba, Ibaraki 305-0801, Japan}
\affiliation{CLASSE,
Cornell University, Ithaca, NY 14850, USA}

\begin{abstract}
The buildup of low-energy electrons has been shown to affect
the performance of a wide variety of particle accelerators.  Of
particular concern is the persistence of the cloud between beam bunch
passages, which can impose limitations on the stability of operation
at high beam current.  We have obtained measurements of
long-lived electron clouds trapped in the field of a quadrupole magnet
in a positron storage ring, with lifetimes much longer than the
revolution period.  Based on modeling, we estimate that about 7\% of
the electrons in the cloud generated by a 20-bunch train of 5.3~{\gev}
positrons with 16-ns spacing and $1.3\times10^{11}$ population survive
longer than 2.3~{${\mu}$s} in a quadrupole field of gradient
7.4~T/m. We have observed a non-monotonic dependence of the trapping
effect on the bunch spacing.  The effect of a witness bunch on the
measured signal provides direct evidence for the existence of trapped
electrons.  The witness bunch is also observed to clear the cloud,
demonstrating its effectiveness as a mitigation technique.
\end{abstract}


\maketitle

\section{Introduction}


Electron cloud buildup has been observed in many accelerators since
the 1960s~\cite{ECLOUD12:Wed0830}.  Adverse consequences of electron
cloud buildup include emittance growth, beam instabilities, and excess heat
load to cryogenic systems.


Positron storage rings for which electron clouds have been an
important factor in the design and performance include KEKB in
Japan~\cite{PRL75:1526} and PEP-II in the USA~\cite{PAC97:2V015}.
Proton accelerators affected by electron clouds include the Los
Alamos Proton Storage Ring (PSR) in the USA~\cite{PRSTAB11:010101}, 
CERN's Proton Synchrotron (PS), Super Proton Synchrotron (SPS)
and Large Hadron Collider (LHC)~\cite{ECLOUD12:Wed0900}.  At the LHC,
electron cloud has been observed to affect 
the cryogenic heat load~\cite{CERN:ATS2013:009}.


Electron cloud buildup is a major concern for accelerator upgrade programs
and for the design of future accelerators.  Electron cloud
considerations have driven the design of the SuperKEKB collider~\cite{JVSTA30:031602}
and the positron damping ring for the proposed
International Linear Collider (ILC)~\cite{PRSTAB17:031002}.  
The LHC luminosity upgrade is contingent on
reducing the bunch spacing to 25~ns~\cite{IPAC11:TUYA02}; at this
bunch spacing, severe electron cloud buildup has been observed, such
that this bunch pattern has been used for beam scrubbing
runs~\cite{ECLOUD12:Wed0900}. The success of the upgrade is likely to
be contingent on limiting electron cloud buildup.


Considerable work has been done on the development of electron cloud
mitigation techniques.  At KEKB and PEP-II, solenoidal magnetic field
windings were installed on the beam-pipes. For SuperKEKB, solenoidal windings are
used in field-free regions, while TiN coatings and antechambers are
included in quadrupole magnets, where solenoidal windings cannot be
used.  Carbon coatings for the dipole magnet vacuum chambers in the
SPS are under study at CERN~\cite{IPAC11:TUPS028}.


The electron cloud is observed to build up during the passage of a
train of closely-spaced bunches, imposing restrictions on the operational 
bunch charge
and train length.  In field-free regions, gaps between
trains allow the electron cloud to dissipate.  In regions of magnetic field, however, 
cloud electrons can become trapped over long periods of time.  
Since trapped electrons can interact with the beam over many turns, they
have the potential for more severe effects.


Electron cloud trapping has been studied experimentally and
via simulation.  Trapping of electrons oscillating around a 70-m-long
proton bunch in the LANL PSR storage ring has been observed.~\cite{PRSTAB11:010101}.  
At LBNL, electrons were observed to be
trapped in the fields of an ion beam and accelerator elements, and
measurements of the time dependence of electron cloud buildup were
carried out~\cite{PRL97:054801}.  Estimates of long-lived electron cloud
buildup at the LHC and consequences for vacuum chamber heat load have
been presented in Ref.~\cite{PRL93:014801}.  More recently, heat load
in the final-focus quadrupoles of the LHC has been attributed to
electron cloud buildup~\cite{UNapoli2014:PHD:GIadarola}.  Simulations
were used to study electron trapping in quadrupole and sextupole
magnets for the parameters of the KEKB positron
ring~\cite{PRE66:036502}, as well as for the Cornell Electron Storage
Ring (CESR) and the ILC positron damping ring~\cite{ECLOUD10:MOD05}.
Prior to the measurements presented here, no experimental study of 
electron trapping in a positron storage ring has been available to validate 
modeling efforts.


A principal goal of the Cornell Electron Storage Ring Test Accelerator
program~\cite{ICFABDNL50:11to33} is to investigate performance
limitations in future high-energy low-emittance rings.  These studies
include measurements of electron cloud buildup caused by
synchrotron-radiation-induced photoemission on the surface of the
vacuum chamber. The CESR ring stores positron and electron beams of
energy 1.8~{\gev} to 5.3~{\gev}, arranged in bunches spaced in
intervals of 4 or 14~ns, with bunch populations ranging up to
$1.6\times10^{11}$.  A variety of detectors sensitive to cloud
electrons incident on the vacuum chamber wall have been used to
study cloud buildup~\cite{NIMA760:86to97,PRSTAB17:061001,IPAC13:MOPWA072,NIMA749:42to46,ECLOUD12:Fri1240}.


The potential for undesired consequences to accelerator
performance motivated the study of electron trapping in
the {\cesrta} electron cloud research program.  
In this paper, we report on the measurement of electron
trapping in a quadrupole magnet over a 2.3~{$\mu$}s time interval 
between bunch train passages. Our demonstration of cloud trapping
is based on two observations: first, the revolution-averaged electron flux arriving
at the vacuum chamber wall during the passage of a ten-bunch train of
positrons is greater when a second such bunch train immediately
follows, showing that cloud is present at the time of arrival of the
first ten-bunch train; second, inserting a single positron bunch over
a broad time range centered halfway around the ring reduces the
observed flux of electrons at the wall during the train passage,
showing that trapped electrons were cleared by the intermediate bunch.
It is noteworthy that
beam-free intervals in the ring are ineffective at clearing the
electrons, since the trapping mechanism is not contingent upon the
beam potential as was the case at the PSR.

\section{Time-Resolving Electron Detector}
\begin{figure}[htbp]
\includegraphics*[width=0.82\columnwidth]{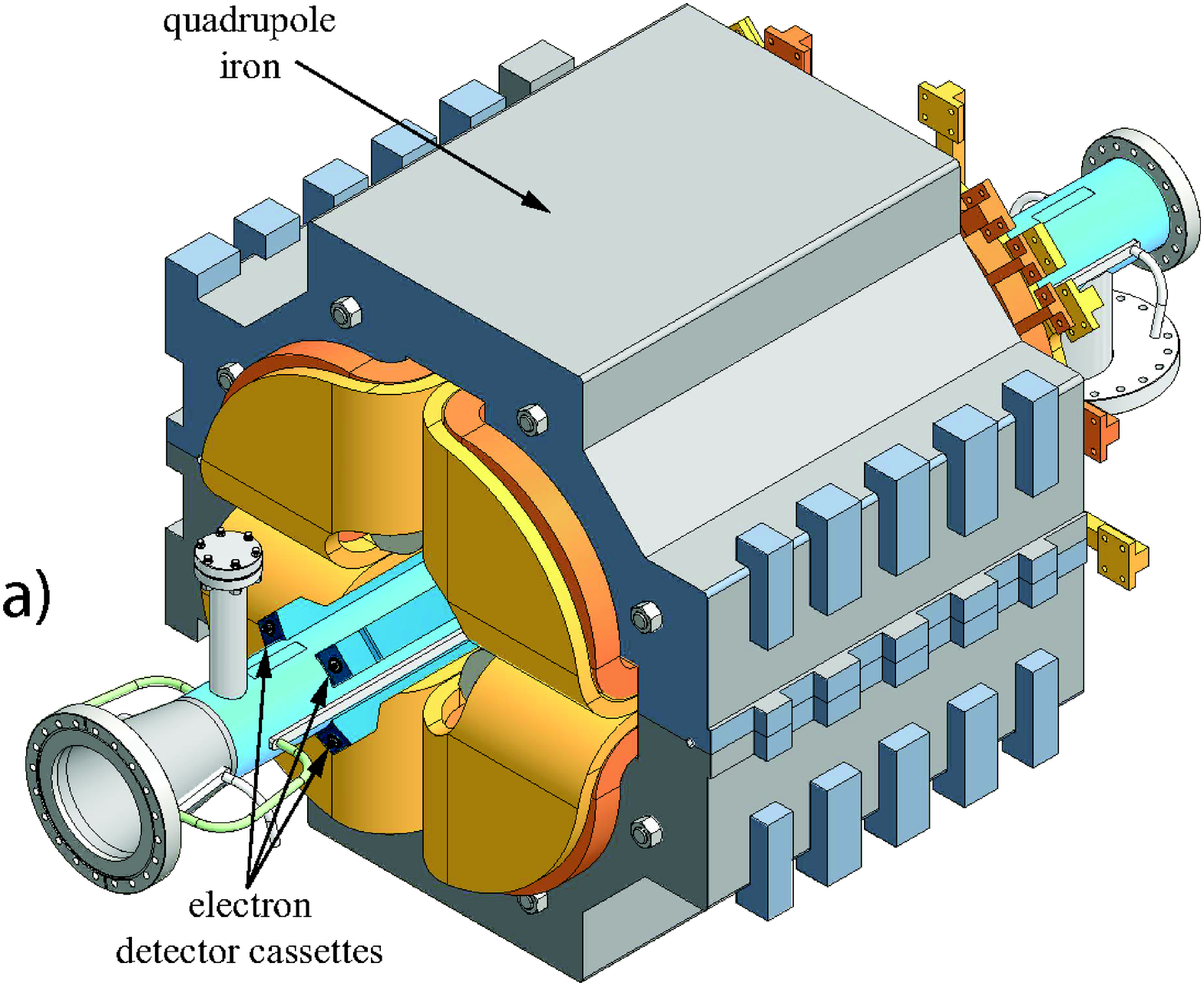}
\vskip 2mm
\includegraphics*[width=0.9\columnwidth]{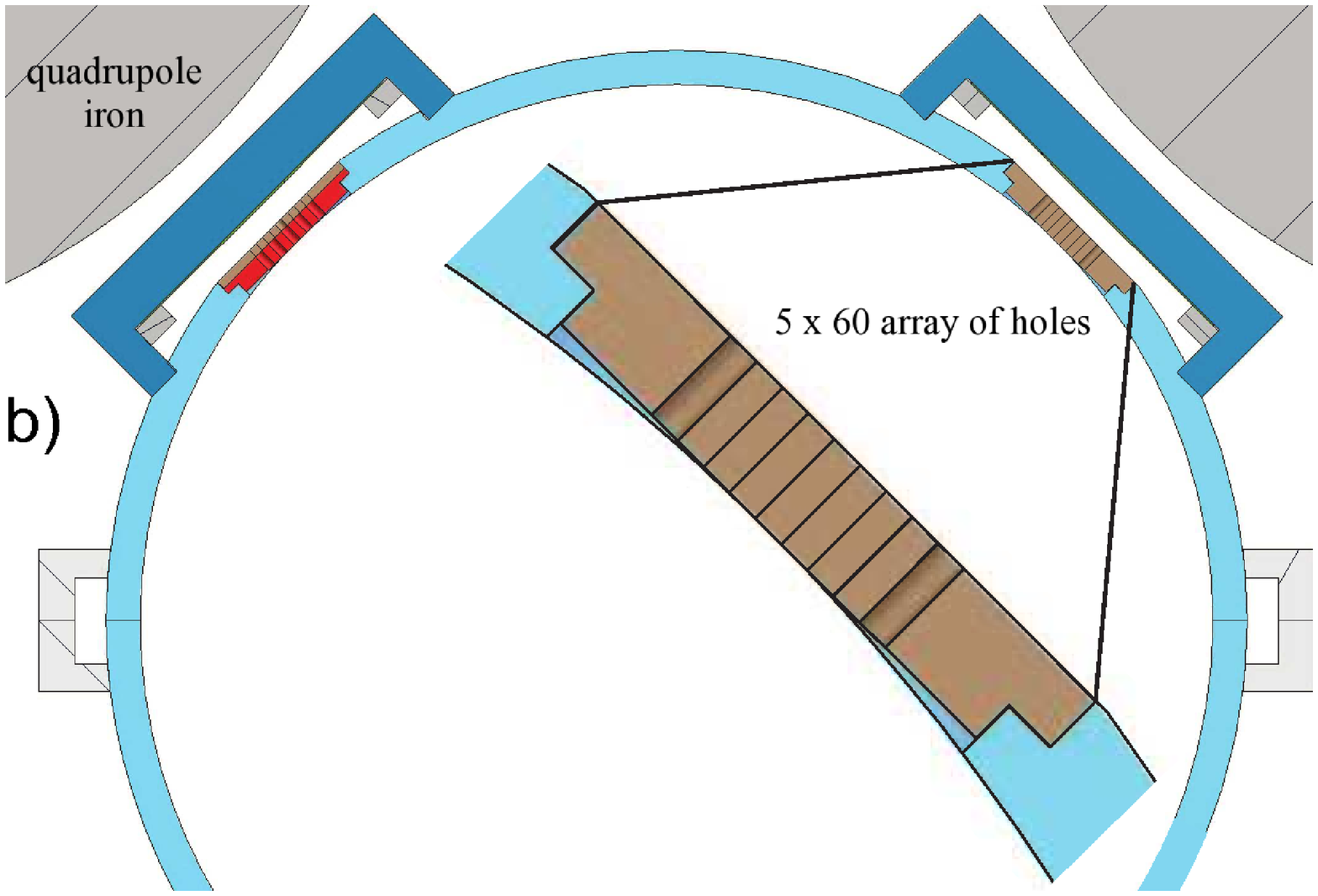}
\vskip 3mm
\includegraphics*[width=0.9\columnwidth]{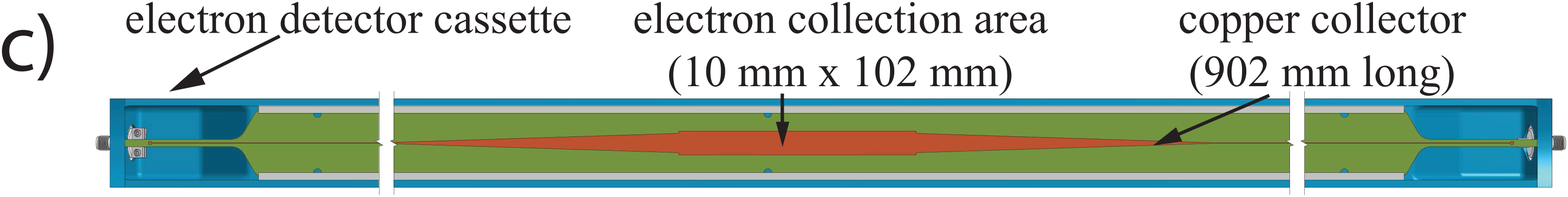}
\vskip -2mm
\includegraphics*[width=0.9\columnwidth]{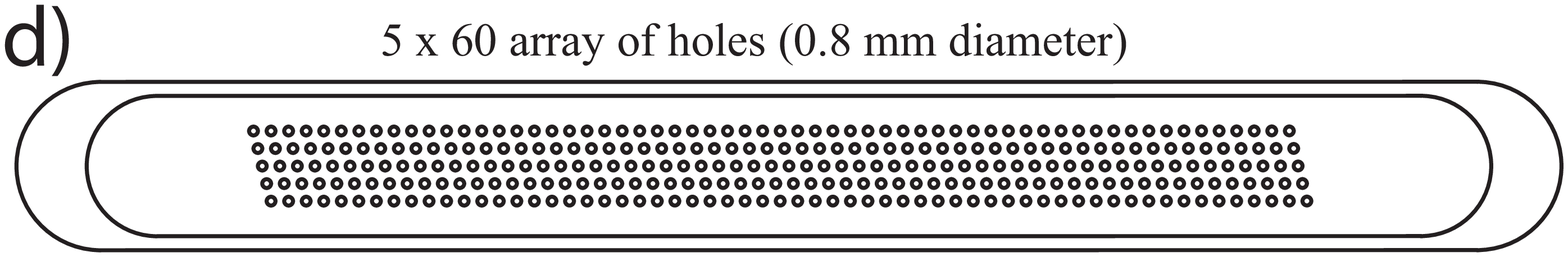}
   \caption{a)~Vacuum chamber equipped with electron detectors in the 
              quadrupole magnet.
            b)~Arrangement of two detectors in front of the magnet poles as 
               seen from the positron arrival direction.
            c)~Geometry of the copper electrode biased at 50~V to 
               collect electrons entering through the pattern of holes 
               in the beam-pipe shown in d). 
               The rectangular region of the collector and
               the pattern of holes are each about 10 cm long.}
   \label{fig:qspu-detector}
\end{figure}

\begin{figure}[htbp]
\includegraphics*[width=0.9\columnwidth]{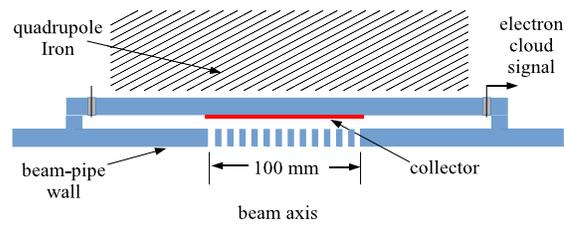}
\caption{Schematic cross section of the electron detector, which is 
         located near the longitudinal center of the quadrupole magnet.
         The holes in the beam-pipe wall allow cloud electrons to reach the collector.}
   \label{fig:schematic}
\end{figure}

Time-resolving electron detectors have provided
detailed information on local cloud formation, allowing the independent
characterization of photoelectron and secondary electron production 
mechanisms~\cite{NIMA749:42to46,ECLOUD12:Fri1240}.
We have installed shielded detectors in
a cylindrical stainless steel vacuum chamber 
of inner diameter 95.5~mm inside a 60-cm-long quadrupole magnet,
as shown in Fig.~\ref{fig:qspu-detector}a. One detector was located longitudinally
near one end of the iron yoke in order to measure electron cloud buildup in the 
fringe field.
In the following, we refer exclusively to measurements obtained from the
detector positioned
in the longitudinal center of the magnet and located 
in azimuth at 45 degrees from the horizontal mid-plane toward the inside of the ring, 
as shown in Fig.~\ref{fig:qspu-detector}b. 
Electrons are collected on the 10-mm-wide copper trace (Fig.~\ref{fig:qspu-detector}c)
which tapers to a transmission line using the grounded copper on the other side of 
the 0.12-mm-thick Kapton sheet. The total length of the trace including the 
10-mm-wide, 102-mm-long rectangular central region is 907 mm.
The pattern of $5\times60$  parallel \mbox{0.8-mm-diameter} holes 
shown in Fig.~\ref{fig:qspu-detector}d
allows passage
of cloud electrons through the beam-pipe to the collector.
The chosen hole diameter gives a depth-to-diameter ratio of 3:1
in order to shield the detector from the RF power radiated by
the 18-mm-long positron bunches~\cite{PEP:253}.
The hole pattern is 7.1~mm wide and 94.4~mm long.
Figure~\ref{fig:schematic} shows a schematic view of the beam-pipe, hole pattern
and detector arrangement.

The collector is biased at $+50$~V relative to the vacuum chamber
in order to prevent secondary electrons from leaving the collector surface.
The AC-coupled front-end readout electronics consists 
of two Mini-Circuits ZFL-500 broadband amplifiers
with $50~{\Omega}$ input impedance and a total gain of 40~dB.
Oscilloscope traces are digitized to 8-bit accuracy in 1000 time bins, 
typically 0.5 or 1.0~ns wide, averaging over 8000 beam-synchronous triggers.
The direct beam-induced signal from the residual transmission
of high-frequency RF power through the shielding holes results in a damped 
ringing in the raw oscilloscope signals. 
All signals depicted in the figures below
show the result of applying a 13-MHz low-pass digital
post-processing filter which suppresses this noise by an order of magnitude. 

\begin{figure}[b]
   \includegraphics*[width=\columnwidth]{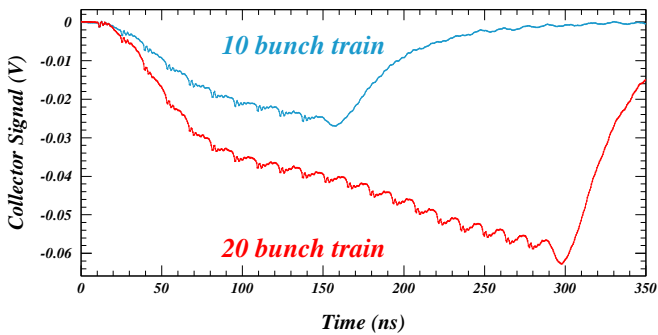}
   \caption{Electron detector signals recorded for  10- and 20-bunch trains of
     5.3~{\gev} positrons for an average bunch population of 
     $1.3\times10^{11}$. 
     The enhanced signal
     during the first 10 bunches of a 20-bunch train relative to that for the 10-bunch 
     train shows that electrons were trapped during the entire 2.3~{{$\mu$}s} interval 
     prior to the return of the bunch train.}
   \label{fig:qspu-signals}
\end{figure}

Figure~\ref{fig:qspu-signals} shows the filtered signals for
10- and 20-bunch trains of 5.3~{\gev} positrons. 
The bunches have rms sizes of 1.8~mm horizontally and 0.08~mm vertically.
The average bunch population is $1.3\times10^{11}$.
The bunch spacing
is 14~ns and the bunch-to-bunch population is uniform to a few percent. 
The quadrupole field gradient is 7.4~T/m, horizontally focusing.

\begin{figure}[tbp]
   \includegraphics*[width=\columnwidth]{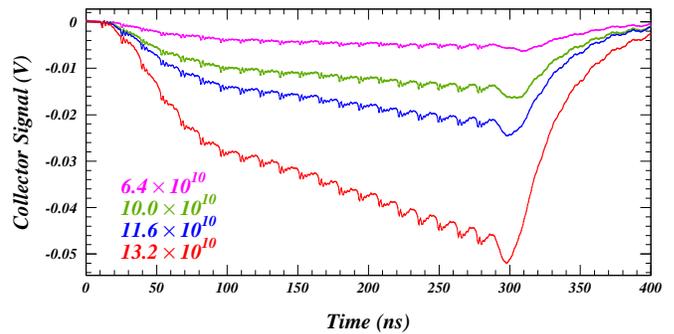}
   \caption{Dependence of the signals on bunch population for 20-bunch trains with 14 ns spacing.
            The dependence is strongly nonlinear, the signal amplitude increasing by an order of
            magnitude for a factor of two increase in bunch population.}
   \label{fig:bunchcurrent}
\end{figure}

The larger signal during the first 10 bunches of the 20-bunch train
relative to that for the 10-bunch train shows the presence of cloud
prior to the arrival of the train. 
One can deduce that electrons remain trapped at least as long as the 2.3~{{$\mu$}s} beam-free interval
prior to the return of the bunch train. 
The decrease in cloud buildup rate following the first 6 bunches
indicates that a subset of trapped electrons which can contribute 
signal has become depleted at that time. In spite of this clearing of the trapped reservoir
of electrons, the signal does not return to the level of the 10-bunch signal, showing that the
additional cloud seeded by the long-term trapping is self-sustaining.
The signal depends strongly on the bunch population,
decreasing by an order of magnitude as the bunch population decreases by a factor of two
from $1.3 \times 10^{11}$ to $6.4 \times 10^{11}$, as shown in Fig.~\ref{fig:bunchcurrent}.

\begin{figure}[tbp]
   \includegraphics*[width=\columnwidth]{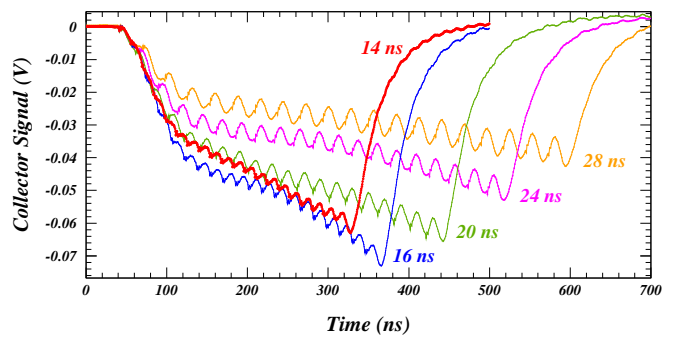}
   \caption{Comparison of signals obtained from 20-bunch trains with spacing 
            14, 16, 20, 24 and 28~ns.
            The increase in signal for the 16-ns spacing relative to the 14-ns spacing 
            shows that long-term cloud electron trapping can be enhanced by an 
            unfortunate choice of bunch spacing.}
   \label{fig:spacing}
\end{figure}

The dependence of trapping on the bunch spacing is shown in Fig.~\ref{fig:spacing}
for a bunch population of about $1.3 \times 10^{11}$.
The decrease with increasing bunch spacing can be understood in terms of an overall
decrease in cloud buildup. However, the enhancement of the signal at 16-ns spacing 
relative to the signal for 14-ns spacing shows that when electron trapping is of concern,
care must be taken in the choice of bunch spacing.

\begin{figure}[tbp]
   \includegraphics*[width=\columnwidth]{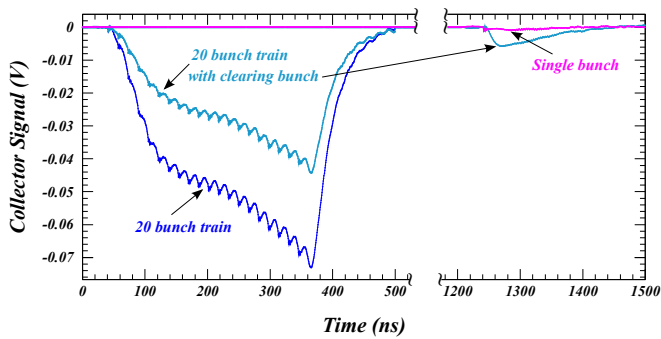}
\vskip -3mm
   \caption{Effect of an intermediate clearing bunch following about 900~ns after the end of
            a 20-bunch train for the case of 16-ns spacing. 
            The difference in magnitude
            between the signals at 1250~ns
            is directly sensitive to the trapped electrons produced by the 20-bunch train.}
   \label{fig:clearing}
\end{figure}

We have investigated the effectiveness of an intermediate bunch as a mechanism
for clearing the trapped cloud. Figure~\ref{fig:clearing}
shows the three 
signals obtained from 1) a 20-bunch train, 2) a 20-bunch train with a clearing bunch following about 900~ns after the end of the train, and 3) a single bunch. 
The single-bunch signal is plotted
to coincide with the signal from the clearing bunch for the purpose of comparison. 
The clearing bunch accelerates trapped cloud 
electrons into the detector, and thus provides direct evidence 
for the trapped cloud. In addition, the reduced signal from
the 20-bunch train when the clearing bunch is present shows the effectiveness of such a 
mitigation technique.
We verified that the clearing effectiveness is independent of the delay of the clearing bunch over a
range of {$\pm$}500~ns. The full clearing effect was achieved when the clearing bunch population reached
about 20\% of the average population of the bunches in the train.

\section{Trapping Mechanism}
The long-term trapping of electrons in nonuniform
fields such as quadrupole fields can be understood 
in terms of an adiabatic magnetic moment $\mu$ given by
\begin{equation}
\mu =  \frac{m v_{\perp}^2}{2 B},
\end{equation}
where $m$ is the mass of the electron, $B$ is the magnetic
field magnitude,
and $v_{\perp}$ is the velocity component perpendicular
to the magnetic field vector (see, for example, Ref.~\cite{RJGoldston1995:IntPlasPhys}).
This quantity remains invariant as long as $\frac{{dB}}{B} \ll 1$
during the cyclotron motion, or, equivalently, 
\begin{equation}
\Gamma = \frac{|\nabla B|r_{\rm c}}{B}  \ll 1,
\end{equation}
where $r_{\rm c}$ is the cyclotron radius.
Combining the conditions of conservation of magnetic moment
and conservation of energy, one can specify a ``velocity-space
loss cone'' angle, $\Theta_{\rm LC}$, which defines the trapping condition.
A  particle moving from a region of lower
field to a region of higher field reverses its path
if the velocity components perpendicular and parallel to the magnetic
field at the starting position, denoted by $v^{\rm in}_{\perp}$ and
$v^{\rm in}_{\parallel}$ respectively, are related such that
\begin{equation}
\frac{v^{\rm in}_{\parallel}}{v^{\rm in}_{\perp}} \le  \left(\frac{B_{\rm bd}}{B_{\rm in}} - 1\right)^{1/2}. 
\end{equation}
Here $B_{\rm in}$ is the magnetic field magnitude at the start point, 
and $B_{\rm bd}$ is the magnitude along the field line at the 
boundary beyond which the particle is lost. 
If the above relationship is satisfied, the particle reaches
a point where the parallel velocity goes to zero, and the particle
reverses its path along the field line. 
In a quadrupole magnetic field, the trapped particle is confined 
between two such mirror points located along a field line 
symmetric about either the horizontal or the vertical axis. While the particle
mirrors between the pair of points, it drifts in the
longitudinal direction until it reaches the fringe region of the
quadrupole, where it can escape~\cite{NAPAC13:TUPAC13}. This drift is caused by a nonzero
gradient and curvature in the magnetic field, often referred to
as the ``grad B'' and ``curvature'' drift respectively. 
For a 7.4~T/m field gradient, the longitudinal 
drift over the duration of one CESR beam revolution is significant only when 
the electron energy is of the order of 1~{\kev}. The energy distribution
obtained from the cloud build-up modeling described below indicate that 
less than 3\% of the electrons have energies exceeding 1~{\kev}.        

\begin{figure}[tb]
   \centering
   \includegraphics*[width=0.97\columnwidth]{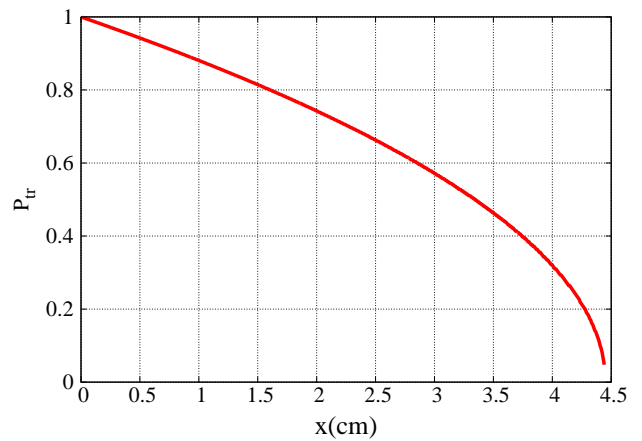}
   \caption{Cosine of the loss-cone angle,  $P_{\rm tr}$, versus 
            horizontal position in the mid-plane of 
            the vacuum chamber. The trapping probability increases
            toward the center of the chamber as long as the adiabaticity
            condition is satisfied.}
   \label{fig:cos_losscone_angle}
\end{figure}

The cosine of the loss cone angle represents the fractional
solid angle in velocity space within which a particle remains confined.
Thus, for a localized distribution of isotropic velocities, it represents the 
probability of confinement at that point.
It can be expressed as 
\begin{equation}
P_{\rm tr} = \cos{\Theta_{\rm LC}} = \left(1 - \frac{B_{\rm in}}{B_{\rm bd}}\right)^{1/2}
\end{equation}
and is shown in Fig.~\ref{fig:cos_losscone_angle} as a function of horizontal position $x$
along the mid-plane of the vacuum chamber.
The probability of confinement decreases 
with $x$, the distance from the beam, provided $\Gamma \ll 1$.
The adiabatic condition can be expressed as 
\begin{equation}
\Gamma = \frac{\sqrt{2mE_{\perp}/e}}{Kx^2} \ll 1, 
\end{equation}
where $e$ is the electron charge, $K$ is the quadrupole field gradient 
and $E_{\perp}$ the kinetic energy corresponding to the velocity component
perpendicular to the magnetic field.  For electrons in a quadrupole with field gradient
\mbox{$K = 7.4$~T/m}, $\Gamma$ reduces to
\begin{equation}
  \Gamma = 4.6 \times 10^{-3} \frac{\sqrt{E_{\perp}/{\rm eV}}}{(x/{\rm cm})^2}.
\label{eq:gamma}
\end{equation}
For comparison, the beam kick produced by a bunch 
carrying $1.3 \times 10^{11}$ positrons on an electron at the 
vacuum chamber wall is 60~eV in the impulse approximation~\cite{LHC:ProjNote:97}, 
easily satisfying the trapping condition. 
On the other hand, an electron with a horizontal momentum
of 40~keV/c located 1~cm from the beam  in the horizontal mid-plane
is likely to hit the chamber wall.

\section{Numerical Modeling of Electron Cloud Buildup}
We have employed a particle-in-cell, time-sliced
cloud buildup modeling code~\cite{PRSTAB5:121002} to improve our understanding
of the electron trapping mechanism and the observed signals. 
The code includes simulation algorithms 
for photoelectron generation,
macroparticle tracking in the 2D electrostatic fields of the beam and 
the cloud, and 3D tracking in a variety of ambient magnetic fields, as well as for a detailed
model of the interaction of cloud electrons with the vacuum chamber surface~\cite{PRSTAB5:124404}.
  
The code has been supplemented with response functions for the  time-resolving
electron detectors~\cite{IPAC14:TUPRI034}. As a function
of incident angle and energy, a fraction of
the macroparticle charge hitting the wall in the region of the detector contributes to the 
modeled signal.
The fraction is derived from an analytic calculation of the hole acceptance for the case of a 
magnetic field
parallel to the hole axis. 
For an arbitrary magnetic field strength, the acceptance of 
the holes
is derived by relating the incident kinetic energy and angle to the 
cyclotron radius
and the wall traversal time, i.e. the fractional number of cyclotron revolutions performed 
in the wall. 
Thus the 
acceptance at high field extends to grazing angles of incidence when the cyclotron radius
is smaller than the hole radius.

The amplitude of the modeled signal was found to be very sensitive to the assumed secondary 
emission
yield, increasing by an order of magnitude as the peak secondary yield was increased from 
1.4 to 1.9.
The measured signal amplitude was reproduced with values for the peak secondary yield and
elastic yield of 1.4 and 0.5, respectively. 

The model shows the signal to be generated predominantly by electrons originally produced 
on the field lines entering the detector, i.e. from a narrow surface region in front of the
diametrically opposed pole and from 4-mm-wide regions on the vacuum chamber surface
in front of the other two poles extending
from the middle of the pole toward the detector. These signal macroparticles spiral around
field lines which pass within a few millimeters of the beam. 
The electrons which remain trapped during the 2.3~$\mu$s prior to the train arrival are cleared
out during the first 6 of the 20 bunch passages, reabsorbed either in the detector or the 
vacuum chamber wall. 
The signal also shows that the cloud development proceeds at 
the higher density level following the clearing, since it does not return to the level of the
signal for a 10-bunch train. The trapping results in a sustained higher cloud density even after the trapped electrons have been removed. 

\begin{figure}[tbp]
   \centering
   \includegraphics*[width=\columnwidth]{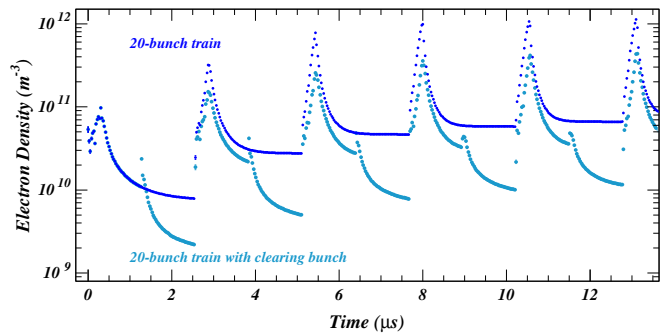}
   \caption{Results for the beam-pipe averaged cloud density from the numerical model 
            of electron buildup
            for the case of the 5.3~{\gev} 20-bunch train of positrons with \mbox{16-ns} 
            spacing shown in Fig.~\ref{fig:clearing}, 
            with and without an intermediate clearing bunch.}
   \label{fig:ecdensity}
\end{figure}
Figure~\ref{fig:ecdensity}
shows the modeled electron cloud density averaged 
over the test volume of the 
cylindrical vacuum chamber for the case of a 20-bunch train
of positrons with average bunch population $1.3\times10^{11}$, with and without an intermediate 
clearing bunch of the same population. 
The peak density in the absence of the clearing bunch reaches $1.1\times10^{12}$~m$^{-3}$ 
after three turns, about 7\% of which is trapped until the train returns. The clearing bunch reduces the
trapped cloud density by about a factor of four.

\begin{figure}[tbp]
   \centering
   \includegraphics*[width=0.9\columnwidth]{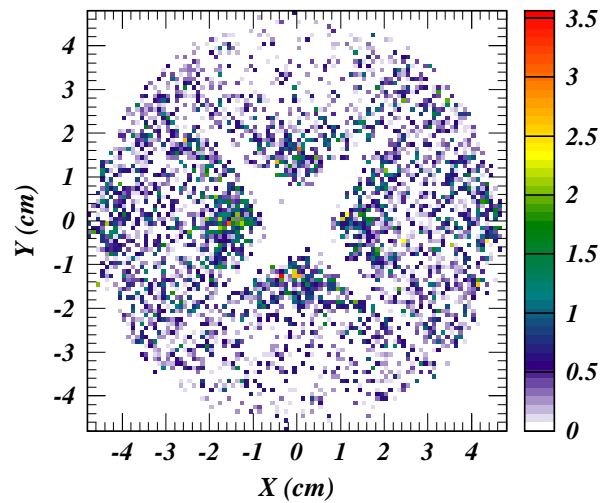}
   \caption{The modeled transverse distribution of the trapped cloud shown
            at the end of the first beam revolution. The color scale ranges up
            to a maximum of $3.5\times10^6$ electrons/bin.}
   \label{fig:ecsnapshot}
\end{figure}

The modeled transverse distribution of the cloud trapped in the quadrupole magnet 
at a time immediately preceding the return of the train 
is shown in Fig.~\ref{fig:ecsnapshot}.
The trapped electrons are
concentrated in four quadrants near the beam outside of a central depletion zone
of 2~cm diameter, consistent with the trapping probability distribution shown
in Fig.~\ref{fig:cos_losscone_angle} and the non-adiabaticity 
in the central and diagonal regions.
The median energy of the trapped electrons is about 50~{\ev}. 

\section{Summary}
Our measurements with a time-resolving electron detector
located in a quadrupole magnetic field have provided
comparisons of signals from
10- and 20-bunch trains of positrons
which show clear evidence for electron
trapping during the entire 2.3~{${\mu}$s} time interval prior to the return of the bunch train.
Modeling tuned to the recorded signals indicates that 
approximately 7\% of the cloud generated by a 5.3~{\gev} train of 20 bunches,
each carrying $1.3\times10^{11}$ positrons, remains trapped. 
The measurements show a non-monotonic dependence on bunch spacing. 
The clearing effect of an intermediate bunch
has been measured and successfully modeled, showing the trapped cloud can be reduced
by a factor of four by such a clearing bunch.
This characteristic of a quadrupole magnetic field to concentrate electrons
near the beam raises concerns for storage rings with positively charged beams,
since those electrons can be attracted into the beam.
Such measurements quantifying electron trapping in quadrupole magnets provide
information useful for the development of simulation codes which serve to predict
electron cloud phenomena in future accelerators and to aid in the
design of mitigation techniques.
\linebreak

\section{Acknowledgments}
We wish to thank Joe Calvey, Gerry Dugan, Miguel Furman and David Rubin for useful discussions.
This work is supported by the US National Science Foundation contracts
\mbox{PHY-0734867}, \mbox{PHY-1002467}, and \mbox{PHY-1068662}, by the US Department of Energy
contract \mbox{DE-FC02-08ER41538} and by the Japan/US Cooperation Program.
\linebreak


\bibliographystyle{medium}
\bibliography{BibListAll_local}
\end{document}